# Transfer Learning Application of Self-supervised Learning in ARPES

Sandy Adhitia Ekahana[1,†,*], Genta Indra Winata[2,†], Y. Soh[1], Gabriel Aeppli[1,3,4], Radovic Milan[1], Ming Shi[1]

*1Paul Scherrer Institute, Forschungstrasse 111, Villigen, 5232, Switzerland*
*2The Hong Kong University of Science and Technology, Clear Water Bay, Kowloon, Hong Kong SAR*
*3Institute of Physics, Ecole Polytechnique Fédérale de Lausanne (EPFL), CH-1015 Lausanne, Switzerland*
*4Laboratory for Solid State Physics, ETH Zurich, Zürich, Switzerland*

[†] Equal contribution

*To whom correspondence should be addressed. Email: sandy.ekahana@psi.ch

**Abstract**

Recent development in angle-resolved photoemission spectroscopy (ARPES) technique involves spatially resolving samples while maintaining the high-resolution feature of momentum space. This development easily expands the data size and its complexity for data analysis, where one of it is to label similar dispersion cuts and map them spatially. In this work, we demonstrate that the recent development in representational learning (self-supervised learning) model combined with k-means clustering can help automate that part of data analysis and save precious time, albeit with low performance. Finally, we introduce a few-shot learning (k-nearest neighbour or kNN) in representational space where we selectively choose one (k=1) image reference for each known label and subsequently label the rest of the data with respect to the nearest reference image. This last approach demonstrates the strength of the self-supervised learning to automate the image analysis in ARPES in particular and can be generalized into any science data analysis that heavily involves image data.

Angle resolved photoemission spectroscopy is a powerful tool to visualize the electronic bandstructure, which has been used mostly in the context of condensed matter physics. The technique evolves stepwise with root back to the photoelectric effect first discovered by Hertz in 1887 [1] and explained theoretically by Einstein in his 1905 annus mirabilis papers [2] (Figure 1(a)). The development of magnetic spectrometer allows the photoelectron first to be resolved in energy (Figure 1(b)) allowing the technique like x-ray photoelectron spectroscopy (XPS) to emerge and flourish as an important tool to analyse any material component from their core level fingerprint [3]. In the 1990s, commercial double-pass cylindrical mirror analyser (CMA) was typically used to resolve the kinetic energy providing better resolution for the experiment [4]. Moving towards the Fermi level of the material, this 1-dimensional (1D) scan (Energy vs photoelectron intensity (I) axis) reveals the density of states (DOS) of the valence band. The second development is based on the fact that each of the 1D detector can be fine-tuned to only accept the electron coming out from a small range of spherical angle allowing us to have an angle-resolved measurement (angle ($\phi$), energy axes vs I) [5-7] (Figure 1(c)) with a typical angular resolution of 1°. Subsequently, these detectors can be placed in an array to make it a 2D analyser (angle ($\phi$), energy axes vs I) simultaneously measuring photoelectrons from a range of angular position [8, 9] with typical angular resolution of ~0.1°; commonly named as an analyser slit. This angular axis allows to resolve the DOS in momentum space direction, i.e. one angle at a particular energy corresponds to a unique momentum axis position. The subsequent development is causing to expand the dimension of the analysis in the perpendicular angle direction, making it a 3D data set (angle 1 ($\theta$), angle 2 ($\phi$), energy axes vs I) (Figure 1(d)), by either moving the 2D analyser slit or by rotating the sample itself [10-13]. This allows us to explore the momentum space in two dimensions, i.e., $(k_x, k_y, E)$ axis. The remaining momentum direction perpendicular to the surface or $k_z$ direction can also be explored by varying the incoming photon energy ($h\nu$) completing the picture of the bandstructure in solids [14, 15]. The time component ($t$) can also be investigated by using an ultrafast technique [16, 17], but we will leave out this additional complexity for the remainder of this paper.

Meanwhile, the introduction of a micrometer size (and smaller) beam spot, with laser or synchrotron-based photon source, expands the technique capability by resolving the data from the 2D analyser spatially. This basically means that we can resolve different bandstructures (angle ($\phi$), energy axes vs I) spatially, forming now 4D data sets of ($x$, $y$, angle ($\phi$), energy axes vs I) (Figure 1(e)). This development marks the exponential increase of the data volume as more information is now being taken as it grows quadratically in general by area, let alone the inclusion of the other omitted "dimension" such as the angle 2 ($\theta$), perpendicular momentum ($k_z$), and the time component ($t$). In most of the so-called micro ($\mu$) or nano (n) - ARPES measurement, the original interest in mapping out the real space domain is done by directly plotting the integrated intensity of the 2D analyser as a function of position i.e. plotting ($x$, $y$ axes vs I), where most of the domain information is contained. For example, the work in reference [18] shows the plot of the integrated intensity over the dispersion cut (angle ($\phi$), energy axes vs I) onto the real space, visualizing domains where different layers of graphene thin film lie. Meanwhile, a synchrotron-based measurement can also trace the core-level or any band lying deep below the Fermi surface to distinguish different sample environments experienced by the element in interest, e.g., areas with different termination [19]. However, there are cases where the core level is not available, e.g., for laser-based ARPES or simply the differences observed are only embedded in the bandstructure as these references show [20, 21].

Labelling each bandstructure observed (angle ($\phi$), energy axes vs I) at a different position ($x$, $y$) can be a tedious task, yet an important one, especially in the context of the experiment. This task can be even more challenging and urgent especially if we are also running against the decay time of the sample surface or simply because of the limited time for the experiment. Plotting the integrated intensity (including intensity normalization on each bandstructure picture) can be a preliminary step to draw and usually is the first-to-go technique. Subsequently, one may try to find a unique region of interest (ROI) that can define a representative bandstructure from the other, yet this approach differs case by case and can take up the precious time during the measurement. Therefore, it is ultimately favorable to have an automated procedure to help the labeling/clustering of these different bandstructures so we can focus more on the physics problem.

Meanwhile, there have been attempts to implement technique available in the machine learning field on scientific experiments like scanning tunnelling microscopy [22-25] and atomic force microscopy [26, 27], where the majority of them are simply a supervised model that needs manual labelling to begin with. There are also similar attempts in ARPES for various purposes. For example, a deep layer of convolutional neural network (ConvNet) is trained to denoise ARPES data [28]. Afterwards, there are efforts to obtain how the bandstructure calculation based on the ARPES data [29, 30], where reference [30] provides additional feature of obtaining the result even through a noisy data (simulated noise). There is also work in automation of spatial domain assignment with a smaller subset of data over a predetermined area, where subsequent position measurement is calculated with Gaussian process to give the possible highest amount of information [31]. Our work here is in the same direction as the reference in [31] which is to automate domain assignment. However, there is a stark difference in our approach that we use representation learning from self-supervised models to represent our ARPES images.

In the context of computer vision, we need to distinguish the term "to classify" and "to cluster". The term classify usually refers to a supervised labelling where there is a definite set of labels these bandstructures can be labeled into. For example, we have a clear pre-defined label of "with-gap" and "no-gap" bandstructures where each ARPES cut can be labeled into. Meanwhile, clustering refers to the same-ness, like-ness, or affinity to a set of standard images to which bandstructures are compared; different image references will create their own different clusters. From ARPES perspective, the final product of "to cluster" and "to classify" can be identical, which is a list of numbers indicating the group this bandstructure belongs to, yet the methods to produce the list are technically different.

Naturally, one may resort to the automated labelling technique called supervised learning, e.g., convolutional neural network (ConvNet) [32] in this case and come up with a trained neural network model that can classify the given bandstructure into the well-defined label. The ConvNet is a well-known and robust machine learning model to learn image representation by learning from examples. Despite this, the ConvNet model may lack generality in the low-resource setting, when the model is trained with a very limited number of data and we need to train the model from scratch. Meanwhile, the amount of data from each measurement may not be sufficient for a neural model to be trained properly, despite the many data enhancement procedures that can be done [30].

**Supervised Learning in ARPES**

We may argue that any well-known pretrained deep neural network, like ResNet50 model which is pretrained with ImageNet dataset, can be used in the model initialisation, where subsequent training with our own training dataset will be done; the processes are described in Figure 2 and the details about the dataset used is described in supporting information section B (SI.B). In short, we run evaluation using k-fold cross validation, and the performance metrics are calculated by taking the average performance on all folds. In this approach, we set the result in a probabilistic manner for each label (total of four labels) on one image. The performance of this finetuned ResNet50 is evaluated by the metrics, such as accuracy, F1, precision, and recall as shown in Table I; refer to Appendix for definition of each metric element. This approach clearly demonstrates some usefulness as the accuracy and the F1 is relatively high, given the training data is relatively few. Thus, this positive result suggests that the possibility to leverage pre-trained models for ARPES data in supervised learning scenarios. In Figure 2, we display how all data are subsequently predicted by finetuning ResNet50 model, where the group probability of each realspace position is shown. We can see that the result of the ResNet50 prediction can be used as a guide to the eye to investigate different domains in the experimental context. However, it should be restated that this approach lacks generality as the fine-tuned ResNet50 model can only be used for the $Fe_3Sn_2$ data; a new sample needs a new re-training which implies the need of a new set of labeled data where this very problem is something we aim to avoid. Following this, the retraining itself can be expensive computationally (involving GPU) and not something favorable to be performed during the ARPES measurement itself as it also costs us valuable time.

| Method | Fold | Accuracy | F1 | Recall | Precision |
|---|---|---|---|---|---|
| ResNet50 | 1 | 90.24 | 73.99 | 67.65 | 89.13 |
| | 2 | 93.06 | 81.31 | 75.43 | 94.20 |
| | 3 | 90.67 | 77.38 | 72.74 | 86.27 |
| | 4 | 89.15 | 73.95 | 71.24 | 82.86 |
| | 5 | 88.89 | 77.68 | 73.33 | 83.45 |
| | AVG | 90.40 | 76.86 | 72.08 | 87.18 |

Table I – Results of finetuning pre-trained ResNet50 on $Fe_3Sn_2$ 4D ARPES data.

**Unsupervised Clustering in ARPES**

Another trivial attempt in the automation can be made by an unsupervised clustering method, such as k-means [33], where the data can be clustered according to their affinity to each other; usually, the Euclidean distance of the feature dimension is used to define the affinity. The typical data processing done before applying this procedure is to flatten the picture, i.e., convert the 2D array of the image into a 1D array and perform the k-means clustering in this "hyperdimensional" array. However, the curse of dimensionality, as coined by Richard Bellman [34], might play a role where this many-dimensions vector effectively hides the feature that needs to be captured. For this issue, one may do the common dimension reduction technique such as principal component analysis (PCA) [35] or a t-distributed stochastic neighbour embedding (t-SNE) [36] to reduce the number of dimensions and subsequently perform the k-means clustering on this reduced dimension space. This approach may help solve the curse of dimension problem, but Table II shows that the performance metric is still comparable when the image data is simply flattened and clustered (method = "none" in Table II). In any case, some kind of "flattening" is still done on the picture where the object of interest is locked into some position in the array. This will greatly affect the result as the feature of interest needs to be on the same array pointer, and some data preprocessing needs to be done to keep the features in place before flattening.

| Method | Dimension Reduction | Accuracy | F1 | Recall | Precision |
|---|---|---|---|---|---|
| DINO | no | 50.79 | 46.59 | 61.73 | 49.95 |
| SwAV | | **61.25** | **59.97** | 68.74 | **59.91** |
| MoCo | | 53.90 | 50.28 | 65.99 | 51.96 |
| BYOL | | 57.05 | 57.58 | **74.40** | 58.24 |
| None | | 53.42 | 54.61 | 72.12 | 59.14 |
| DINO | | 43.80 | 40.61 | 62.23 | 45.55 |
| SwAV | | 49.28 | 40.60 | 60.59 | 47.09 |
| MoCo | yes (t-SNE 2-d) | 50.10 | 41.92 | 59.34 | 43.81 |
| BYOL | | 46.83 | 43.56 | 62.50 | 47.86 |
| None | | **58.97** | **48.97** | **67.48** | **48.72** |

Table II – Results of k-means clustering using pre-trained model and None as bare-image, with and without dimension reduction by t-SNE.

**Self-Supervised Learning in ARPES**

Recently, pre-trained self-supervised models are utilized to generate high-dimensional representation of data, where instead of comparing the input data directly, the model will map the input data onto a representational space on which further training can be done, or simply using them for clustering [37]; the input data can be speech, text, or images (refer to supporting information for details). In short, these models learn the feature from the input data in an self-supervised way, which means the training objective is generated without using annotated human labels, but using pretext tasks. In the context of natural science experiment, there are already attempts to apply some form of self-supervised learning model, for example in the training of self-supervised model built for single cell image analysis [38], training of self-supervised model to denoise tomography data (Noise2inverse) [39, 40], analysis of electrocardiography (ECG) database with a general purpose self-supervised model [41], quantifying hidden feature in single-molecule charge transport data with transfer learning [42], etc. Meanwhile, the field of computer science has already introduced several general-purpose self-supervised models for computer vision, such as MoCo [43], SimClr [44], BYOL [45], SwAV [46], and DINO [47] where ARPES can take advantages on through *transfer learning*, i.e., utilise a pre-trained model, that was trained using self-supervision fashion on large image dataset (e.g., ImageNet [48]). The summary of the procedure is described in supporting information and the clustering pipeline is shown in Figure 3. Table II shows the result of k-means clustering performance metrics done on the representation space of DINO, SwAV, MoCo, BYOL, with and without t-SNE dimension reduction. We can see that the performance is worse than the ResNet50 supervised performance (Table I) and unfortunately is even comparable without using a pre-trained self-supervised model (method = "none") on the image; dimension reduction also offers no significant help in this approach. This result may imply that the feature of ARPES image is not captured well in the knowledge extracted from the ImageNet database; this invites the need for a collective ARPES database for training purposes and also a novel objective task of self-supervised model for ARPES in specific or any scientific method which relies heavily on images. It can also mean that ARPES images from $Fe_3Sn_2$ dataset are not well separated in the representation space; the representation learning space on images may be well spread, yet they do not make well separate islands on which k-means clustering can be of any use.

As a current solution, we introduce a k-nearest neighbour (kNN) procedure (the pipeline is shown in Figure 4), where instead of direct clustering on the representation space, we take k-numbers of images as reference on which the rest of the images can be compared with. The references in here are the images with a known label set by the expert. The input images and the references are first converted into the representation space by passing to pre-trained self-supervised model accordingly and the Euclidean distances of the converted images are measured with respect to the converted references. The label is then determined by the closest reference. In our case of $Fe_3Sn_2$, we have 4 labels and thus k-numbers of images on each label. In the case of $k = 1$, we can see that this approach is equivalent to generating k-means clustering prediction but with a pre-defined centroid (4 pre-defined centroids for case of $Fe_3Sn_2$). This problem can be a quick solution in the case when the images in the representation learning space are not well separated yet they are well spread. As we increase the number of k, we expect the model to be more able to discriminate the given input. The pipeline of this process is drawn in Figure 4.

The summary of the kNN performances are shown in Table III. We can see that as we set $k = 20$ their performances are roughly similar with each other with ranking as follows: DINO, none, BYOL, MoCo, SwAV. This result is expected as the more known labels are given, we are approaching more towards a supervised method like ConvNet above. This result also tells us that different representational technique capture different features of ARPES data differently as their performances vary between techniques. Importantly, we can see that the DINO technique captures ARPES features better than the other technique. Reducing the number of k down to single reference case $k = 1$ also reveals the strength of DINO model [47] (we average 50 different single references for each technique) as the performance still triumphs appreciably as compared to the other techniques.

| Method | k | Accuracy | F1 | Recall | Precision |
|---|---|---|---|---|---|
| DINO | 1 | **80.36** | **70.66** | **80.56** | **70.84** |
| | 5 | **89.16** | **82.09** | **90.62** | **78.73** |
| | 10 | **91.23** | **85.14** | **93.26** | **81.14** |
| | 20 | **92.94** | **87.65** | **95.16** | **83.43** |
| SwAV | 1 | 71.37 | 61.59 | 73.13 | 64.49 |
| | 5 | 82.12 | 74.35 | 85.35 | 72.11 |
| | 10 | 84.62 | 77.56 | 88.62 | 74.13 |
| | 20 | 87.36 | 80.71 | 91.34 | 76.56 |
| MoCo | 1 | 69.74 | 59.25 | 71.19 | 61.36 |
| | 5 | 82.17 | 72.74 | 84.37 | 69.21 |
| | 10 | 85.89 | 77.27 | 87.98 | 72.84 |
| | 20 | 88.32 | 80.51 | 90.48 | 75.71 |
| BYOL | 1 | 76.61 | 65.82 | 75.36 | 68.22 |
| | 5 | 84.68 | 76.75 | 86.72 | 73.96 |
| | 10 | 86.89 | 79.74 | 89.67 | 75.96 |
| | 20 | 88.71 | 82.21 | 92.07 | 77.86 |
| None | 1 | 76.41 | 66.81 | 77.17 | 68.78 |
| | 5 | 83.94 | 76.39 | 88.21 | 72.63 |
| | 10 | 87.00 | 80.03 | 91.38 | 75.35 |
| | 20 | 90.06 | 83.90 | 94.02 | 78.90 |

Table III – Results of the kNN few-shot experiment using pre-trained representation learning.

From the kNN results, we propose the following pipeline to finally solve our problem of automated domain labeling (Figure 5). As the spatial scan begins, we take the first image as our reference $k = 1$ case. Afterward, the subsequent measured images' affinity to the reference is calculated (in representation learning space). For each measured image, the affinity is considered if it has exceeded a certain threshold or not. The threshold itself is an arbitrary value that will be determined as the experiment runs (the expert may intervene during the experiment to decide the value of the threshold). At some point, the Euclidean distance of an image might be relatively further than the others. In this situation, we (as the expert during the measurement) may decide if the new image is regarded as a new label or not, and thus we may set if a new label is needed. Afterward, the experiment again proceeds with the updated number of labels. Finally, the experiment loop will end with a complete set of ARPES dataset and their label as measured with respect to the decided references. The method described here is still not fully automatic but it is a semi-supervised way of automation and is already a great improvement with the help of the learned representation from self-supervised models.

**Conclusion**

We demonstrate that ARPES, which heavily relies on pictorial data analysis, can take advantage of the recent development of representational learning in the computer vision field for help in automation. In this work, we demonstrate the transfer learning application of pre-trained self-supervised models that are utilised on ARPES images in supervised and unsupervised manner. We apply the k-nearest neighbor method, where the affinities of the ARPES images are measured with respect to image reference, and we show that only using a single reference image can achieve decent performance. The kNN proposal presented here is not limited to ARPES image analysis and can be used in other scientific experiments that heavily analyse images and scarce labelled data, given a suitable representation learning method is used. Our work urgently invites the expansion of an ARPES image database on which further development of a machine learning model can be trained. Consequently, it also invites further investigations in an interdisciplinary research collaboration between ARPES and computer vision to find a more suitable representational learning method for ARPES purpose in particular, and another collaboration between natural science and computer science in general, especially in this era of open science.

**Acknowledgement**

We thank Benjamin Bejar Haro from Laboratory for Simulation and Modelling, Paul Scherrer Institute, for the insightful discussion on the k-shot model.

We thank Felix Baumberger, Anna Tamai, and team for the wonderful micro focused laser ARPES setup.

We thank Matthew Watson from i05 Diamond Light Source for the useful discussion in spatial resolved ARPES.

S.A.E acknowledges the support from the European Research Council HERO Synergy grant SYG-18 810451, NCCR-MARVEL funded by Swiss National Science Foundation, the European Union's Horizon 2020 research and innovation programme under the Marie Skłodowska-Curie grant agreement No 701647.

# Appendices

## A. Definitions

|  |  | Predicted Condition | |
|---|---|---|---|
|  | Total Population = P + N | Predicted Positive (PP) = TP + FP | Predicted Negative (PN) = FN + TN |
| Truth condition | Positive (P) = TP + FN | True Positive (TP) | False Negative (FN) |
|  | Negative (N) = FP + TN | False Positive (FP) | True Negative (TN) |

**Accuracy (ACC)**

$$ACC = \frac{TP + TN}{P + N}$$

**F1**

$$F1 = \frac{2TP}{2TP + FP + FN} = \frac{2 \times PRE \times REC}{PRE + REC}$$

**Recall (REC) or True Positive Rate (TPR) or Sensitivity (SEN)**

$$REC = \frac{TP}{P} = \frac{TP}{TP + FN}$$

**Precision (PRE) or Positive Predictive Value (PPV)**

$$PRE = \frac{TP}{PP} = \frac{TP}{TP + FP}$$

From the definition above, imbalance data may create stark differences between REC and PRE as in our $Fe_3Sn_2$ example where this value is reflected in F1. In this case, F1 is a good indicator for the goodness of the model as it "averages" REC and PRE.

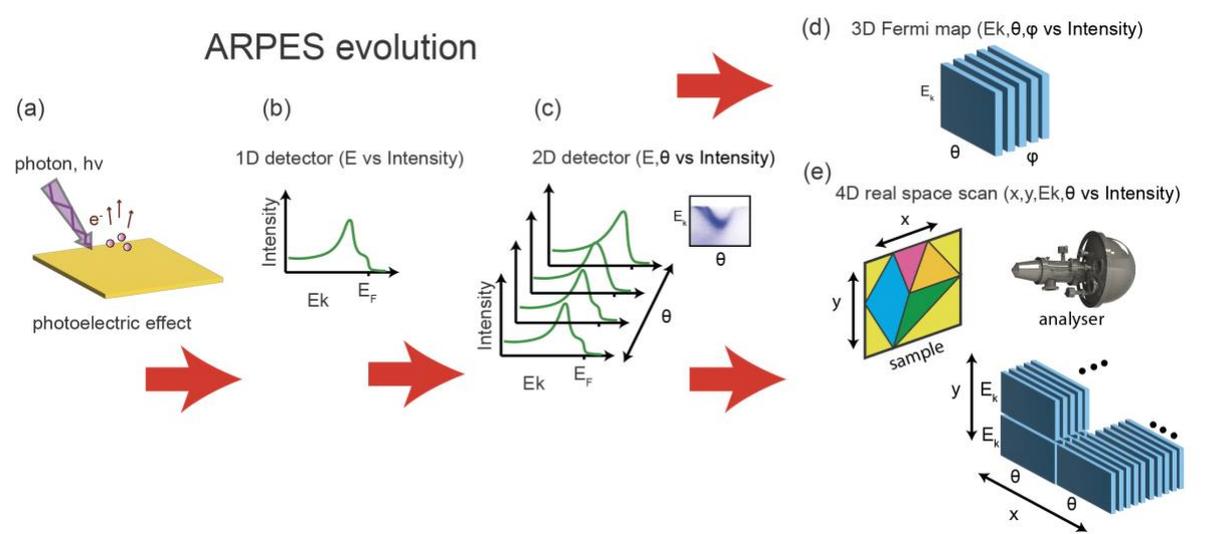

**Figure 1 - Angle Resolved PhotoEmission (ARPES) technique evolution.** (a) ARPES is based on the photo-electric effect where a single electron is ejected out from a sample after one photon illuminating the sample. (b) The photoelectron kinetic energy can be detected with a 1D detector ($E_k$ vs Intensity). (c) Further development of the detector shows a 2D detector where electron from different angle position can be collected simultaneously ($E_k, \theta$ vs Intensity). (d) This 2D detector can be rotated to collect different band dispersion from different angular position ($E_k, \theta, \phi$ vs Intensity) to create a 3D Fermi map. (e) The advent of small beamspot allows a real-space scan creating a 4D data where different band dispersion ($E_k, \theta$ vs Intensity) from different position ($x, y$) are collected i.e. ($x, y, E_k, \theta$ vs Intensity)

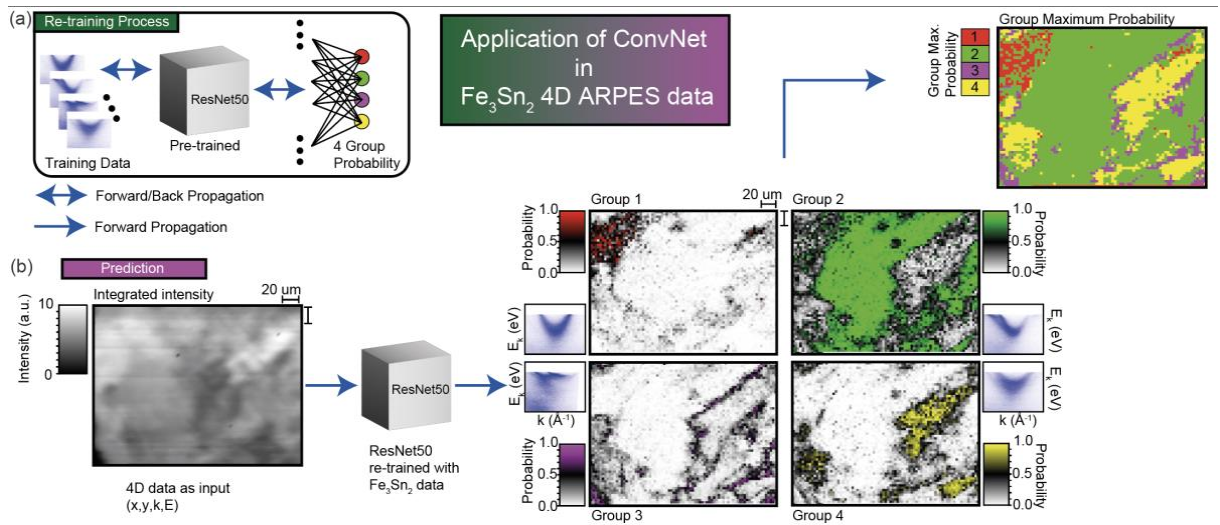

**Figure 2 – Application of supervised learning on Fe$_3$Sn$_2$ ARPES 4D data.** (a) Pre-trained ResNet50 model is re-trained with training dataset (fractions from total data) of labeled Fe$_3$Sn$_2$ band dispersion (4 labels in total as output). (b) The re-trained ResNet50 model is used to predict the rest of the Fe$_3$Sn$_2$ ARPES 4D data with output of probability for each group. We can see that group 2 dominates the data population. The domain picture from the maximum probability label is also shown.

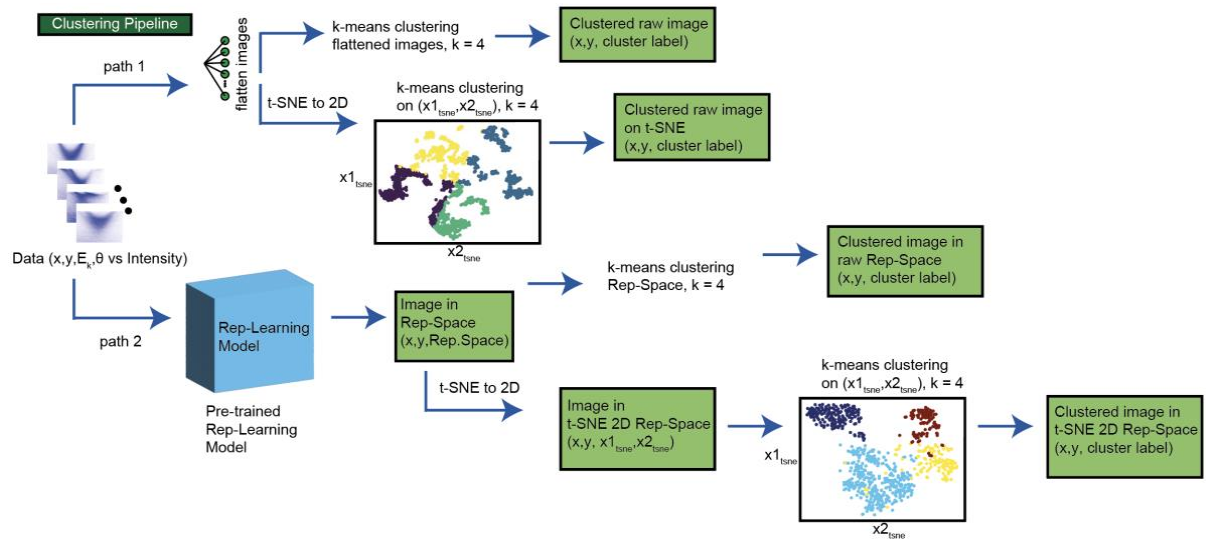

**Figure 3 – Clustering Pipeline.** The raw ARPES images are to be clustered with the k-means algorithm. Path 1 follows the procedure where the raw ARPES images are directly flattened and used for k-means algorithm, with and without dimension reduction (t-SNE onto 2-dimension). Path 2 follows the procedure where additional ARPES image conversion into representational space (rep-space) is done prior to k-means clustering, with and without further dimension reduction of rep-space. The final results of all procedures give us the collection of labels for each input images.

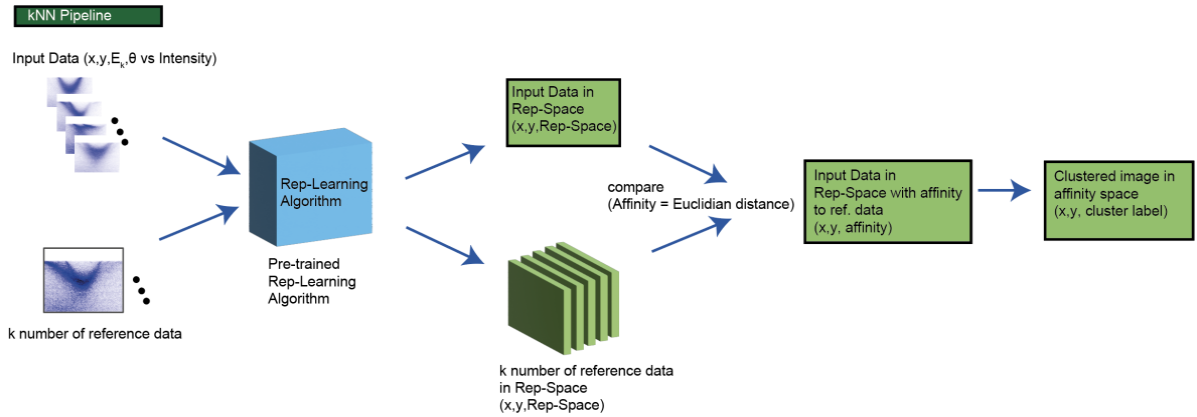

**Figure 4 – k-Nearest Neighbour (kNN) pipeline.** k numbers of images are taken as references for each label where all input data will be compared to (Euclidian-distance-way). The distance calculation is done in representational space where the features of the ARPES data are extracted. Subsequently, the output is the original set of ARPES images with the nearest reference image as their labels. In this work, we show that the minimum of $k = 1$ is enough to be a reference for each label when a self-supervised model is used.

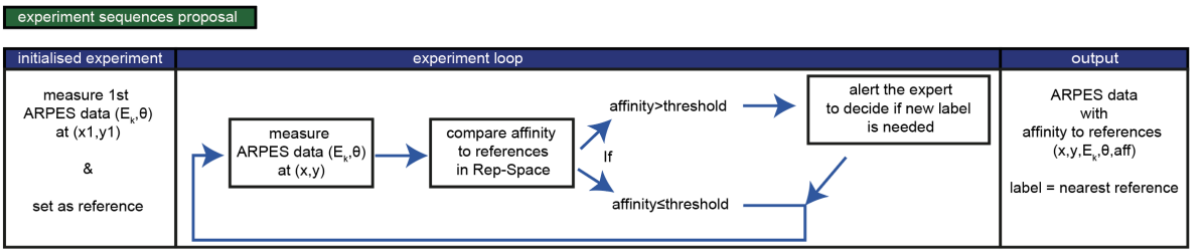

**Figure 5 – Experiment sequences proposal.** For implementing kNN with $k = 1$, we propose the experimental sequences as shown above. In short, as the experiment goes on, the model will alert the expert if an image deviates from the known references where the expert can decide if a new label is needed. The loop continues until the end of the experiment sequence. The output will the 4D ARPES data and also the label on each data point where the label is the closest reference.